\documentclass[%
 reprint,
 amsmath,amssymb,
 prl,
]{revtex4-1}

\usepackage{graphicx}
\usepackage{dcolumn}
\usepackage{bm}
\usepackage{times}
\usepackage{multirow}
\usepackage{epstopdf}
\usepackage{color}
\usepackage[normalem]{ulem}
\usepackage{lineno}

\definecolor{cmntblue}{rgb}{0.0, 0.58, 0.71}
\definecolor{cmntgreen}{rgb}{0.0, 0.42, 0.24}

\usepackage[draft, inline, nomargin]{fixme}
\fxsetup{theme=color, mode=multiuser}
\FXRegisterAuthor{bl}{1}{\color{blue}BL}  
\FXRegisterAuthor{dj}{2}{\color{red}DJ}
\FXRegisterAuthor{nb}{3}{\color{cmntblue}NB}
\FXRegisterAuthor{pm}{4}{\color{cmntgreen}PM}
\FXRegisterAuthor{kh}{5}{\color{green}KH}

\newcommand{\LS}{liquid scintillator (LS) \renewcommand{\LS}{LS}}

\newcommand{\nuebar}{\ensuremath{\overline{\nu}_{e}} }

\newcommand{\NIST}{National Institute of Standards and Technology (NIST) \renewcommand{\NIST}{NIST}}
\newcommand{\NBSR}{National Bureau of Standards Reactor (NBSR) \renewcommand{\NBSR}{NBSR}}
\newcommand{\INL}{Idaho National Laboratory (INL) \renewcommand{\INL}{INL}}
\newcommand{\ATR}{Advanced Test Reactor (ATR) \renewcommand{\ATR}{ATR}}
\newcommand{\ORNL}{Oak Ridge National Laboratory (ORNL) \renewcommand{\ORNL}{ORNL}}
\newcommand{\HFIR}{High Flux Isotope Reactor (HFIR) \renewcommand{\HFIR}{HFIR}}
\newcommand{\LLNL}{Lawrence Livermore National Laboratory (LLNL) \renewcommand{\LLNL}{LLNL}}

\begin{document}

\preprint{APS/123-QED}

\title{First search for short-baseline neutrino oscillations at HFIR with PROSPECT} 
\author{
J.~Ashenfelter$^{p}$,
A.B.~Balantekin$^{m}$,
C.~Baldenegro$^{i}$,
H.R.~Band$^{p}$,
C.D.~Bass$^{f}$,
D.E.~Bergeron$^{g}$,
D.~Berish$^{j}$,
L.J.~Bignell$^{a}$,
N.S.~Bowden$^{e}$,
J.~Bricco$^{n}$,
J.P.~Brodsky$^{e}$,
C.D.~Bryan$^{h}$,
A.~Bykadorova~Telles$^{p}$,
J.J.~Cherwinka$^{n}$,
T.~Classen$^{e}$,
K.~Commeford$^{b}$,
A.J.~Conant$^{c}$,
A.A.~Cox$^{l}$,
D.~Davee$^{o}$,
D.~Dean$^{i}$,
G.~Deichert$^{h}$,
M.V.~Diwan$^{a}$,
M.J.~Dolinski$^{b}$,
A.~Erickson$^{c}$,
M.~Febbraro$^{i}$,
B.T.~Foust$^{p}$,
J.K.~Gaison$^{p}$,
A.~Galindo-Uribarri$^{i,k}$,
C.E.~Gilbert$^{i,k}$,
K.E.~Gilje$^{d}$,
A.~Glenn$^{e}$,
B.W.~Goddard$^{b}$,
B.T.~Hackett$^{i,k}$,
K.~Han$^{p}$,
S.~Hans$^{a}$,
A.B.~Hansell$^{j}$,
K.M.~Heeger$^{p}$,
B.~Heffron$^{i,k}$,
J.~Insler$^{b}$,
D.E.~Jaffe$^{a}$,
X.~Ji$^{a}$,
D.C.~Jones$^{j}$,
K.~Koehler$^{n}$,
O.~Kyzylova$^{b}$,
C.E.~Lane$^{b}$,
T.J.~Langford$^{p}$,
J.~LaRosa$^{g}$,
B.R.~Littlejohn$^{d}$,
F.~Lopez$^{p}$,
X.~Lu$^{i,k}$,
D.A.~Martinez~Caicedo$^{d}$,
J.T.~Matta$^{i}$,
R.D.~McKeown$^{o}$,
M.P.~Mendenhall$^{e}$,
H.J.~Miller$^{g}$,
J.M.~Minock$^{b}$,
P.E.~Mueller$^{i}$,
H.P.~Mumm$^{g}$,
J.~Napolitano$^{j}$,
R.~Neilson$^{b}$,
J.A.~Nikkel$^{p}$,
D.~Norcini$^{p}$,
S.~Nour$^{g}$,
D.A.~Pushin$^{l}$,
X.~Qian$^{a}$,
E.~Romero-Romero$^{i,k}$,
R.~Rosero$^{a}$,
D.~Sarenac$^{l}$,
B.S.~Seilhan$^{e}$,
R.~Sharma$^{a}$,
P.T.~Surukuchi$^{d}$,
C.~Trinh$^{b}$,
M.A.~Tyra$^{g}$,
R.L.~Varner$^{i}$,
B.~Viren$^{a}$,
J.M.~Wagner$^{b}$,
W.~Wang$^{o}$,
B.~White$^{i}$,
C.~White$^{d}$,
J.~Wilhelmi$^{j}$,
T.~Wise$^{p}$,
H.~Yao$^{o}$,
M.~Yeh$^{a}$,
Y.R.~Yen$^{b}$,
A.~Zhang$^{a}$,
C.~Zhang$^{a}$,
X.~Zhang$^{d}$,
M.~Zhao$^{a}$}
\address{$^{a}$Brookhaven National Laboratory, Upton, NY 11973, USA}
\address{$^{b}$Department of Physics, Drexel University, Philadelphia, PA 19104, USA}
\address{$^{c}$George W.~Woodruff School of Mechanical Engineering, Georgia Institute of Technology, Atlanta, GA 30332, USA}
\address{$^{d}$Department of Physics, Illinois Institute of Technology, Chicago, IL 60616, USA}
\address{$^{e}$Nuclear and Chemical Sciences Division, Lawrence Livermore National Laboratory, Livermore, CA 94550, USA}
\address{$^{f}$Department of Physics, Le Moyne College, Syracuse, NY 13214, USA}
\address{$^{g}$National Institute of Standards and Technology, Gaithersburg, MD 20899, USA}
\address{$^{h}$High Flux Isotope Reactor, Oak Ridge National Laboratory, Oak Ridge, TN 37830, USA}
\address{$^{i}$Physics Division, Oak Ridge National Laboratory, Oak Ridge, TN 37830, USA}
\address{$^{j}$Department of Physics, Temple University, Philadelphia, PA 19122, USA}
\address{$^{k}$Department of Physics and Astronomy, University of Tennessee, Knoxville, TN 37996, USA}
\address{$^{l}$Institute for Quantum Computing and Department of Physics and Astronomy, University of Waterloo, Waterloo, ON N2L 3G1, Canada}
\address{$^{m}$Department of Physics, University of Wisconsin, Madison, Madison, WI 53706, USA}
\address{$^{n}$Physical Sciences Laboratory, University of Wisconsin, Madison, Madison, WI 53706, USA}
\address{$^{o}$Department of Physics, College of William and Mary, Williamsburg, VA 23185, USA}
\address{$^{p}$Wright Laboratory, Department of Physics, Yale University, New Haven, CT 06520, USA}

\collaboration{PROSPECT Collaboration}
\email{prospect.collaboration@gmail.com}

\date{\today}

\begin{abstract}
This Letter reports the first scientific results from the observation of antineutrinos emitted by fission products of $^{235}$U at the High Flux Isotope Reactor. 
PROSPECT, the Precision Reactor Oscillation and Spectrum Experiment, consists of a segmented 4~ton $^6$Li-doped liquid scintillator detector covering a baseline range of 7-9\,m from the reactor and operating under less than 1 meter water equivalent overburden. 
Data collected during 33 live-days of reactor operation at a nominal power of 85\,MW yields 
a detection of 25461\,$\pm$\,283 (stat.) inverse beta decays.  
Observation of reactor antineutrinos can be achieved in PROSPECT at 5$\sigma$ statistical significance within two hours of on-surface reactor-on data-taking.   
A reactor model independent analysis of the inverse beta decay prompt energy spectrum as a function of baseline constrains significant portions of the previously allowed sterile neutrino oscillation parameter space at 95\,\% confidence level and disfavors the best fit of the Reactor Antineutrino Anomaly at 2.2\,$\sigma$ confidence level.  

\end{abstract}

\maketitle


Experiments at nuclear reactors have led to the first direct observation of antineutrinos~\cite{cowan1956}, the discovery of electron antineutrino oscillation~\cite{KamLAND}, and many precise neutrino oscillation parameter measurements~\cite{bib:prl_rate,bib:reno,bib:dc}. 
Nuclear models are used to predict the flux and energy spectrum of electron antineutrinos ($\overline{\nu}_e$) emitted from the decay of 
fission products. 
Absolute \nuebar flux measurements show a $\sim$~6\,\% deficit with respect to recent calculations~\cite{bib:huber,bib:mueller2011}, with this deficit appearing to be dependent on the fuel content of nearby reactors~\cite{bib:prl_evol}.  
The measured spectrum also deviates from model predictions~\cite{bib:prl_rate,reno_bump,dc_bump}.  
It has been suggested that these discrepancies indicate incomplete reactor models or nuclear data~\cite{vogel_review}, oscillation of \nuebar to sterile neutrinos~\cite{bib:mention2011}, or a combination these effects.
A range of experimental~\cite{bib:prl_sterile,bib:neos,danss_osc,bib:neutrino4,stereo_2018}, theoretical~\cite{sonzogni_insights,hayes_shoulder,hayes_shape,neutrons_shoulder,fin_forbidden}, and global analysis~\cite{huber_shoulder,giunti_evol,surukuchi_flux,huber_shoulder,schwetz_global} efforts have sought to understand the origin of these discrepancies.

In a schematic one active plus one sterile neutrino mixing scenario, the oscillation hypothesis predicts reactor \nuebar disappearance due to an eV-scale sterile neutrino described by
\begin{equation}\label{eq:osc}
\begin{aligned}
P_{\rm{dis}} = \sin^22\theta_{14} \sin^2 \left(1.27 \Delta m^2_{41}({\rm eV}^2) \frac{L({\rm m})}{E_\nu ({\rm MeV})}\right),
\end{aligned}
\end{equation}
where $L$ and $E_{\nu}$ are the experimental baselines and neutrino energies, $\Delta$m$^2_{41}$ is the mass squared difference between mass eigenstates, and $\theta_{14}$ is the mixing angle between active and sterile flavor states~\cite{Agashe:2014kda}.   
Widely-cited global fits of this oscillation model to historical neutrino data including reactor flux and spectrum measurements have suggested values of $\Delta$m$^2_{41}$ and $\theta_{14}$ of $\sim~2$~eV$^2$ and $\sim~0.15$, respectively~\cite{bib:mention2011}; we refer to this as the `Reactor Antineutrino Anomaly' oscillation parameter space.  
New experiments seek to unambiguously test this hypothesis via differential measurements of the \nuebar energy spectrum over a range of short ${\cal O}(10)$\,\,m baselines~\cite{danss_osc,bib:neutrino4,stereo_2018,prospect,solid}.  
Such efforts are complicated by the need to perform precision \nuebar measurements in the challenging background environment close to a reactor core and near the Earth's surface with little overburden~\cite{prospect}.  

Using a novel detector concept, PROSPECT is designed to make a reactor-model independent search for short-baseline oscillation and provide a high-precision measurement of the $^{235}$U \nuebar spectrum at a highly-enriched uranium (HEU) reactor.  
This Letter describes the first surface detection of reactor \nuebar by PROSPECT and a model-independent search for sterile neutrino oscillations at the High Flux Isotope Reactor (HFIR) at Oak Ridge National Laboratory.  

PROSPECT consists of a single segmented detector surrounded by a passive shielding package operated at a fixed position near the HFIR core~\cite{prospect_nim, prospect}. 
The cylindrical reactor core (diameter~$=0.435\,$m, height~$=0.508$\,m) uses fuel enriched in $^{235}$U. 
HFIR operates at a fixed power of 85\,MW$_{\textrm{th}}$ for $24$~day cycles, with fresh fuel being used for each cycle. 
A detailed reactor core model incorporating typical fuel and operational data~\cite{Ilas15} indicates that the $^{235}$U fission fraction always remains above 99\,\%.
The PROSPECT detector is deployed in a ground level room adjacent to the water pool containing the HFIR core.  
In this position, the HFIR building provides less than one meter-water-equivalent of vertical concrete overburden, and the HFIR core center is located $\sim~45\,^{\circ}$ below the horizontal from the detector center at a distance of $(7.9\pm 0.1)$\,m.  

The PROSPECT detector is a $\sim~2.0$\,m$\times$1.6\,m$\times$1.2\,m rectangular volume containing $\sim~4$~tons of pulse shape discriminating (PSD) liquid scintillator (LS) loaded with $^6$Li to a mass fraction of 0.1\,\%~\cite{prospect_p20, prospect_p50}.  
Thin specularly reflecting panels divide the LS volume into an $11\times14$ two-dimensional array of 154~optically isolated rectangular segments (14.5\,cm$\times$14.5\,cm$\times$117.6\,cm).  
Hollow plastic support rods 
secure panels in position at segment corners, with row-adjacent segments being vertically offset to create space for the rods outside the active segment volume.
The segment long axis is almost perpendicular ($79\,^{\circ}$) to the vector between the reactor and detector centers.
The LS volume of each segment is viewed by two 5\,inch photomultiplier tubes (PMTs) housed in mineral oil-filled acrylic boxes.  
Thirty-five (42) support rod axes have been instrumented with removable (stationary) radioactive (optical) calibration sources, enabling {\sl in situ} calibration throughout the target volume.  
The detector structure and LS are contained within a rectangular acrylic vessel under a continuous flow of nitrogen cover gas, which is itself housed inside a light-tight aluminum tank. 

PMT signals from collected scintillation light in a segment are recorded using CAEN~V1725 250\,MHz 14-bit waveform digitizer (WFD) modules~\cite{disclaimer}.  
Above-threshold ($\sim$~5\,photoelectron) signals from both PMTs in a single segment are required to trigger zero-suppressed readout of the full detector.  
Trigger rates of roughly 30\,kHz and 5\,kHz are achieved during reactor-on and reactor-off running.  
To avoid ambiguity related to detector re-triggering, analysis cuts actively remove closely-timed triggers, resulting in a dead time of $<$\,2\,\% ($<$\,1\,\%) during reactor-on (-off) periods that is directly determined from data.  

For analysis, PSD, energy, and longitudinal position ($z$) values for particle interactions in a single segment are collected in a \textit{pulse}.  
PSD values for individual PMTs (``tail/total'' ratio of ADC integrals relative to the waveform leading edge) are combined in a weighted average to produce one value for each pulse.  
Pulse energy is determined by summing the ADC integral from each PMT waveform and applying $z$-dependent light collection factors determined from background neutron captures on $^6$Li (denoted nLi). 
Relative pulse arrival times and ADC integral ratios are used to reconstruct $z$. 
Using a 20\,ns coincidence requirement, pulses are grouped into \textit{clusters}.  
Cluster energy, E$_{\textrm{rec}}$, is summed over all contained pulses.  
Cluster $z$-position and segment number, Z$_{\textrm{rec}}$ and S$_{\textrm{rec}}$, are taken from the highest-energy pulse.  
Along with pulse PSD values, these are the primary quantities used in signal selection and  physics analyses.

\begin{figure}[t]
\includegraphics[trim = 0.3cm 0.0cm 1.3cm 0.5cm, clip=true, width=0.49\textwidth]{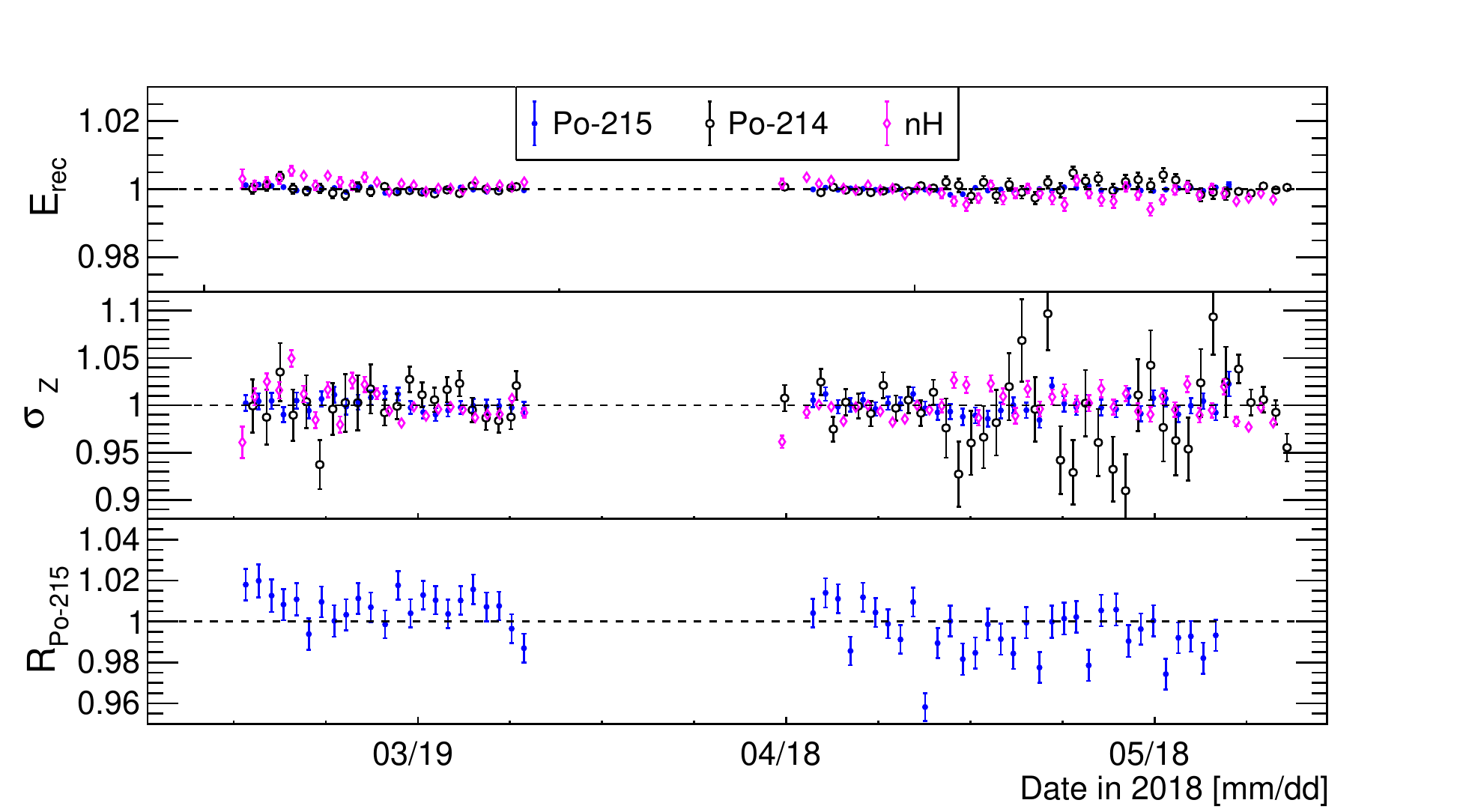}
\includegraphics[trim = 0.3cm 0.0cm 1.3cm 1.0cm, clip=true, width=0.49\textwidth]{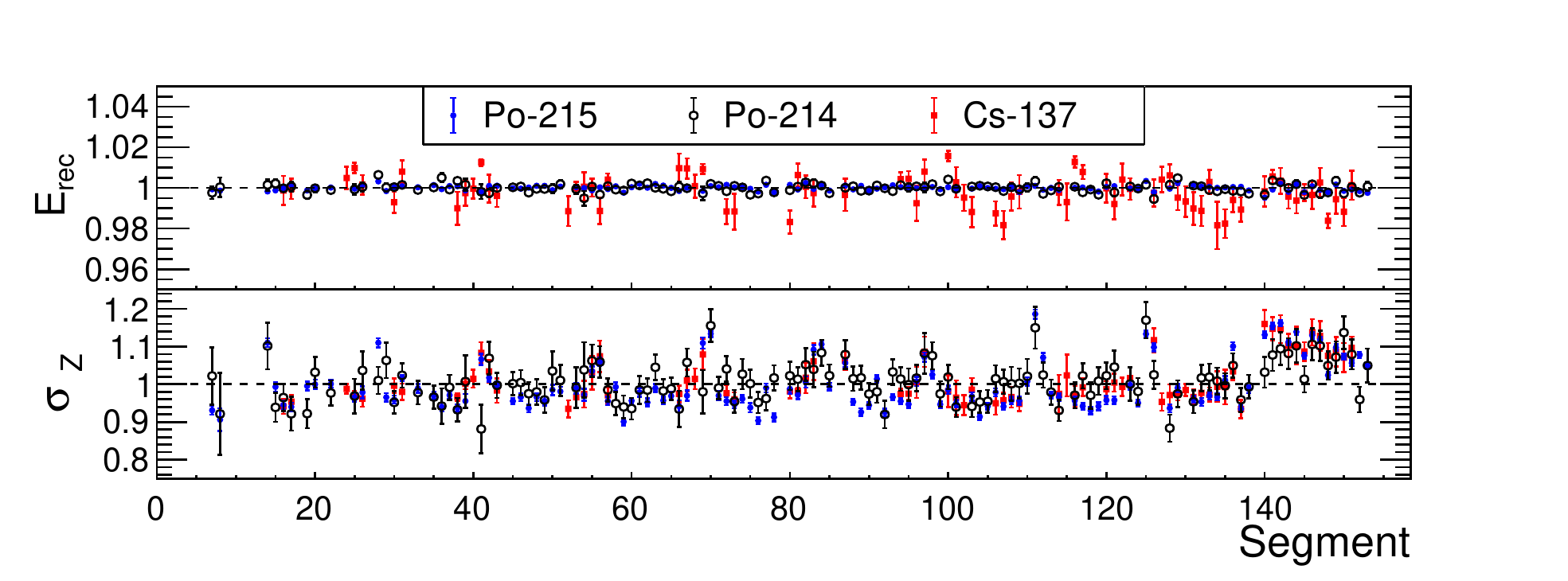}
\caption{Detector time stability and segment-to-segment uniformity in energy (E$_{\textrm{rec}}$), longitudinal position resolution ($\sigma_z$), and detection rate (R).  
Quantities are calculated for $^{214}$Po (black) and $^{215}$Po (blue) alpha decays and neutron-hydrogen captures uniformly distributed throughout the detector (magenta), and for $^{137}$Cs (red) source deployments.  
Reconstructed quantities are described in more detail in the text.  
All quantities are shown relative to the average of all points in the dataset.  
All error bars represent statistical uncertainties.}
\label{fig:response}
\end{figure}

Detector response stability and uniformity are demonstrated via examination of reconstructed physics quantities as a function of time and segment number (Fig.~\ref{fig:response}). 
Sources include high-purity samples of detector-intrinsic ($^{219}$Rn,$^{215}$Po) correlated decays from $^{227}$Ac deliberately dissolved in the LS, ($^{214}$Bi,$^{214}$Po) correlated decays from $^{238}$U, background neutron captures on hydrogen, and $^{137}$Cs source $z$-scans along multiple axes.   
Reconstructed energies ($z$ positions) and energy resolutions ($z$ resolutions) are stable to within $\sim$1\,\% ($\sim$5\,cm) and $\sim$10\,\% ($\sim$10\,\%), respectively, over all times and segments.  
Additionally, the rate of (Rn,Po) events 
is stable to within $\sim~2$\,\%, consistent with the expected 0.7\,\% variation due to the half-life of $^{227}$Ac.

This Letter reports \nuebar measurements based on 33 live-days of reactor-on and 28 of reactor-off data taken between March and May 2018. 
During this data taking period PMTs in 31~segments exhibited intermittent bias current instabilities (19 inside the outer ring of segments, or fiducial volume).  
While this behavior is investigated, segments that at any time exhibited instability are excluded from the analysis.  
This corresponds to a 20\,\% volume reduction (18\,\% in the fiducial volume), in addition to a reduction in detection efficiency for nearby segments as described below.  

PROSPECT detects reactor \nuebar via the inverse beta decay (IBD) interaction, 
$\nuebar + p \rightarrow e^+ + n$, with analysis cuts focused on the selection of a time- and position-correlated prompt positron signal and delayed signal from nLi.  
IBD candidates are selected via the following criteria: 
a prompt cluster of any size with the PSD of all cluster pulses within 3.0$\sigma$ of the gamma-like PSD band mean;
a delayed single segment cluster with $0.46<\textrm{E}_{\textrm{rec}}<0.60$ MeV and PSD more than 3.6$\sigma$ above the gamma-like PSD band mean~\cite{prospect_p50}; 
a coincidence time difference $\Delta$t of (+1,+120)\,$\mu$s;
and a requirement that prompt and delayed clusters lie within identical or horizontally/vertically adjacent S$_{\textrm{rec}}$, with an added $z$-coincidence requirement of 18\,cm and 14\,cm for coincidences in identical or adjacent S$_{\textrm{rec}}$, respectively.  
IBD candidates with the delayed cluster in a (0,+100)\,$\mu$s window around cosmic muon clusters (E$_{\textrm{rec}}>$15\,MeV) or a (-200,+200)\,$\mu$s window around other high-PSD pulses with E$_{\textrm{rec}}>0.25$\,MeV are also rejected.  
These veto criteria result in a well determined inefficiency between 5.5\,\% and 6.9\,\% during this data taking period that varies due to contamination from time-varying $\gamma$-ray backgrounds~\cite{prospect_reactor}.  
Finally, IBD candidates with S$_{\textrm{rec}}$ in the outermost layer of segments or Z$_{\textrm{rec}}$ within 14\,cm of a cell end are rejected.  

The primary backgrounds to the PROSPECT \nuebar measurement are time-correlated signals from cosmogenic neutrons~\cite{prospect} and accidental coincidences of ambient $\gamma$-ray fluxes and nLi captures.  
Accidental coincidence rates during reactor-on and reactor-off periods are calculated with little statistical uncertainty using a $\Delta$t selection of (-12,-2)\,ms.  Cosmogenic background rates and spectra are estimated by applying the IBD selection to reactor-off data.  
The reactor-off correlated event rate is adjusted by $<1$\,\% to account for relative differences in atmospheric pressure, and thus cosmogenic fluxes, between reactor-on and reactor-off datasets~\cite{bib:nucifer}; this factor is determined via measurement and correlation of multiple cosmogenic event classes with local atmospheric pressure measurements~\cite{weather}.  
The resulting reactor-on cosmogenic neutron background prediction is then conservatively assigned a 5\,\% normalization uncertainty.  
Other time-correlated backgrounds 
are expected to contribute $<1$\,\% of the reactor-off sample.  

\begin{figure}[t]
\includegraphics[trim = 0cm 0cm 0cm 0cm, clip=true, width=0.49\textwidth]{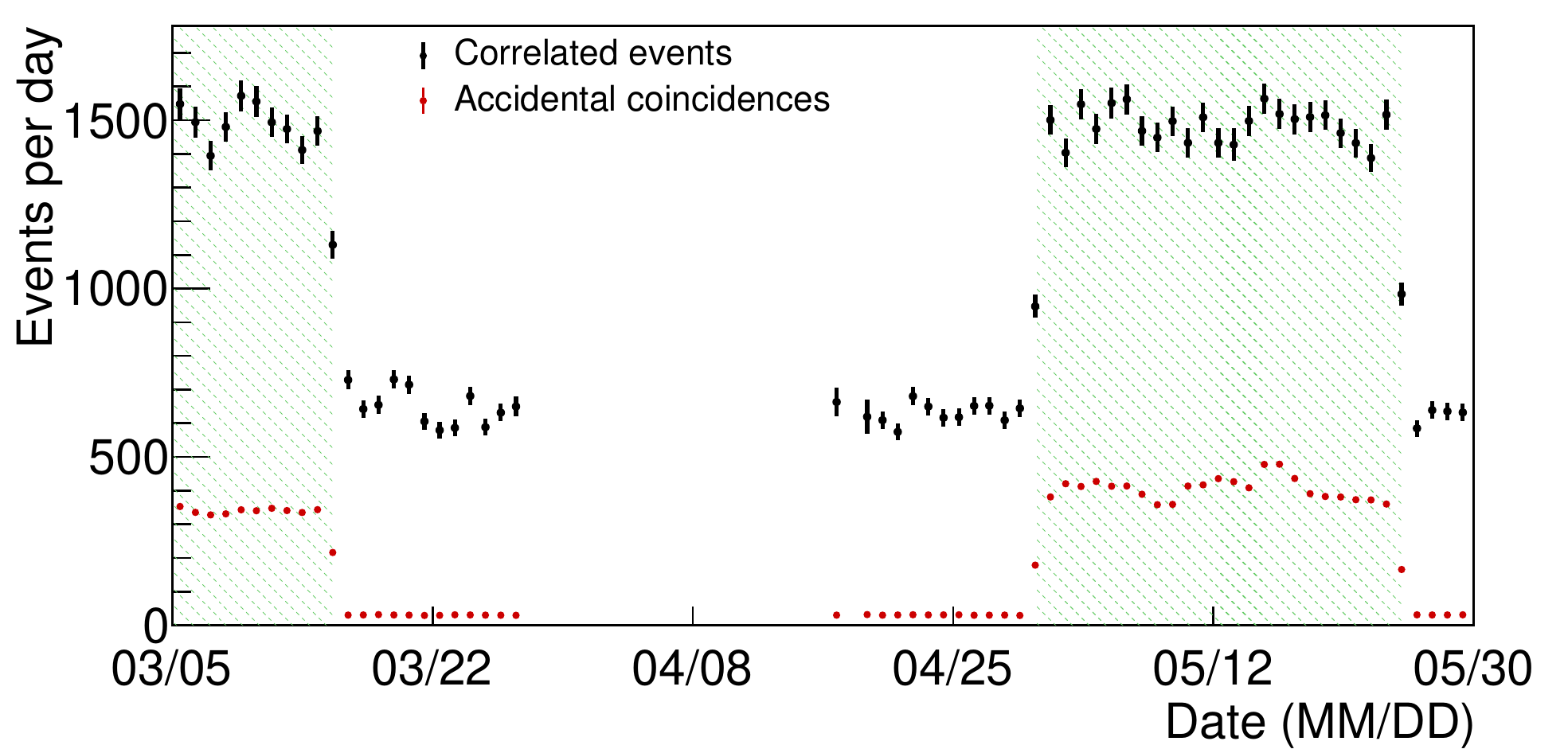}

\vspace{0.15cm}
\includegraphics[trim = 0.2cm 0cm 1.7cm 0.0cm, clip=true, width=0.49\textwidth]{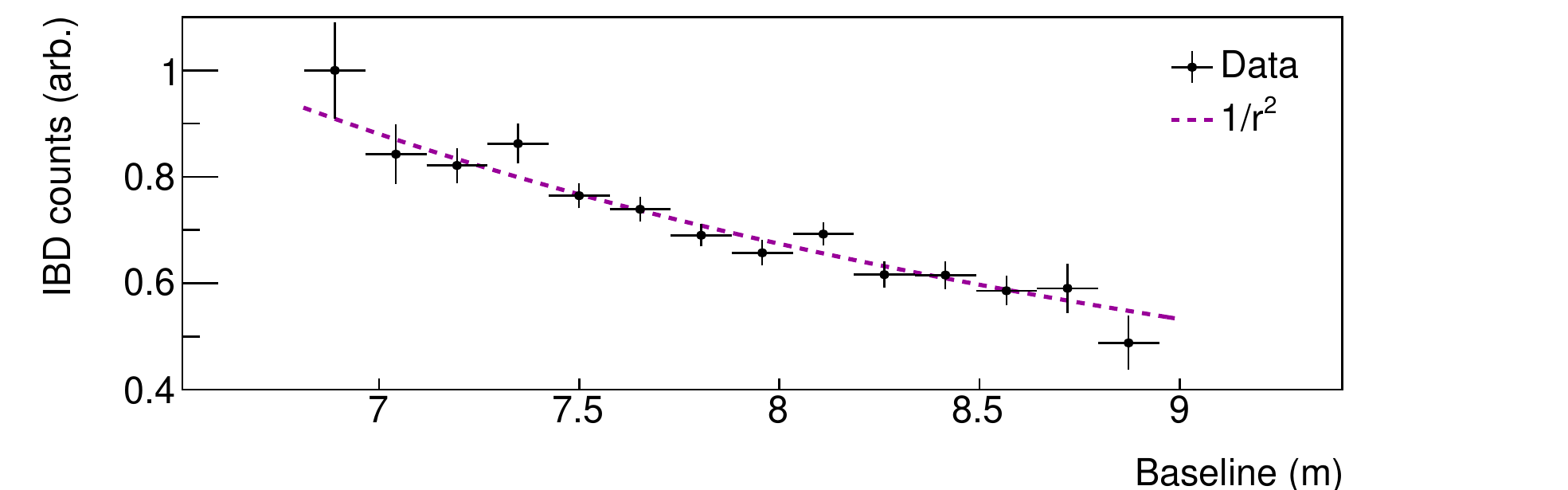}
\caption{Top: Accidentals-subtracted daily IBD-like candidates (black) and calculated accidental coincidences (red).  IBD candidate event rates are corrected for time-dependent variations in detector veto and livetime.  Shaded regions correspond to reactor-on periods.  
The gap in reactor-off data points corresponds to a planned period of detector maintenance and calibration.  Bottom: Normalized background-subtracted IBD event rate versus baseline.  The data is consistent with 1/r$^2$ behavior.  All error bars represent statistical uncertainties.}
\label{fig:selected}
\end{figure}

Between prompt reconstructed energies (E$_{\textrm{rec,p}}$) of 0.8\,MeV and 7.2\,MeV the reactor-on dataset contains 56378 IBD candidates and an estimated $11580\pm 12$ accidental coincidences, yielding $44797\pm 238$ correlated events. 
The corresponding number of correlated background events in the reactor-off data set is $19337\pm 153$. 
Correlated background subtraction yields $25461\pm 283$ detected IBDs (771/day), with a signal-to-background ratio (S/B) of 2.20 and 1.32 for accidental and correlated backgrounds, respectively. 
The correlated event rate for $(0.8<\textrm{E}_{\textrm{rec,p}}<7.2)$\,MeV as a function of time and relative IBD detection rate versus baseline are shown in Fig.~\ref{fig:selected}.  
The difference in the correlated event rate between reactor-off and -on periods indicates a clear detection of IBD events above background.  
The expected 1/r$^2$ variation in IBD rate within the detector is also observed.    
Using the correlated background rate averaged over the entire reactor-off period, the transition to reactor-on operation using the \nuebar signal alone can be identified to 5$\sigma$ statistical significance within 2\,hours.  

\begin{figure}[t]
\includegraphics[trim = 0.0cm 0.0cm 0.0cm 0.0cm, clip=true, width=0.49\textwidth]{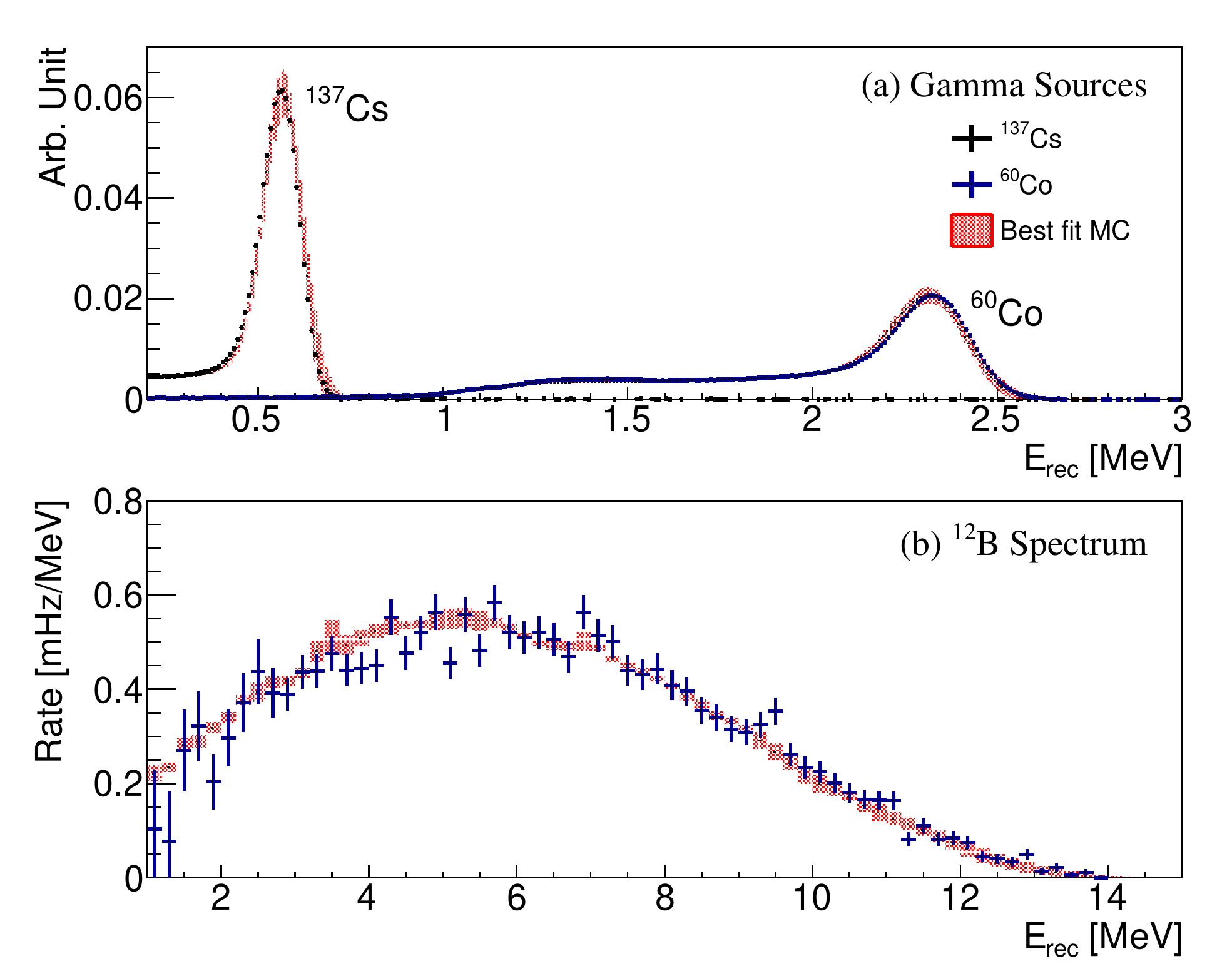}
\caption{a) Measured and best fit simulated E$_{\textrm{rec}}$ spectra for separate $^{137}$Cs and $^{60}$Co $\gamma$-ray calibration runs deployed in the detector center.  b) Observed and best fit MC-predicted reconstructed energy spectra for uniformly distributed beta decays of cosmogenic $^{12}$B.  The red bands represent the energy model uncertainty in the prediction.}
\label{fig:mc_model}
\end{figure}

To perform a differential test of oscillation-induced spectral distortion, an IBD response model is generated for all detector positions using PG4, a GEANT4-based~\cite{bib:Geant4} Monte Carlo (MC) simulation package developed by the collaboration.  
Accurate energy scale, non-linearity, and energy resolution simulation are established via a simultaneous fit to the energy spectra of $^{137}$Cs, $^{22}$Na, and $^{60}$Co center-deployed calibration sources and the $\beta+\gamma$-ray spectrum of cosmogenic $^{12}$B distributed uniformly throughout the detector volume.  
MC data is generated for each calibration dataset in PG4 using an energy response model with two LS non-linearity parameters, 
one photo-statistics resolution parameter, and one absolute energy scale parameter. 
The E$_{\textrm{rec,p}}$ spectra of $^{137}$Cs, $^{60}$Co, and $^{12}$B are shown in Fig.~\ref{fig:mc_model} along with PG4-simulated spectra generated using the best fit 4-parameter set.  
Non-linearities for the best fit model are $\sim$20\,\% over the relevant E$_{\textrm{rec,p}}$ range with a best fit photo-statistics energy resolution of 4.5\,\% at 1\,MeV.  
Model uncertainties, treated as correlated between all segments, are derived by sampling from sets of the 4 model parameters that yield a $\chi^2$ value within 2\,$\sigma$ of the best fit parameter set.  

Accuracy of PG4-reported energy loss is checked using $z$-position scans of a $^{22}$Na $\gamma$-ray source that produces spectral features at $\sim~1.6$\,MeV and $\sim~2.0$\,MeV for detector-edge and detector-center calibration axes, respectively.  
Observed spectrum shifts of up to $\sim~30$\,keV between $z=0$\,cm (segment midpoint) and $z=30$\,cm deployments are correctly reproduced in MC to $\pm$\,10\,keV.  
This $10$\,keV envelope, as well as the 1\,\% time stability of E$_{\textrm{rec}}$ observed for (Rn,Po) and (Bi,Po) are treated as both segment-correlated and segment-uncorrelated energy scale uncertainties.  

Relative detection efficiency variations between segments are modeled with PG4 IBD simulations.   
The largest factor contributing to efficiency non-uniformity is capture of IBD neutrons in segments currently excluded from the IBD selection.  
To understand this effect, data-MC comparisons of IBD candidate prompt-delayed Z$_{\textrm{rec}}$ and S$_{\textrm{rec}}$ coincidence were performed.  
Combined with the previously-mentioned 2\% variation in (Rn,Po) detection rates versus time, this source of uncertainty is conservatively propagated as a 5\,\% segment-uncorrelated IBD rate uncertainty.  

To test for the possible existence of sterile neutrino oscillations, measured prompt energy spectra are compared between different baselines. 
For this purpose, a $\chi^2$ is defined as:
\begin{equation}\label{eq:abs}
\chi^2 = \bm{\Delta}^{\textrm{T}}\textrm{V}_{\textrm{tot}}^{-1}\bm{\Delta}.  
\end{equation}
$\bm{\Delta}$ is a 96-element vector representing the relative agreement between measurement and prediction in 6 position bins and 16 energy bins: 
\begin{equation}\label{eq:delta}
\Delta_{l,e} = M_{l,e}- M_{e}\frac{P_{l,e}}{P_{e}}.  
\end{equation}
In this expression, $M_{l,e}$ and $P_{l,e}$ are the measured and predicted content of the $l^{th}$ position bin and $e^{th}$ $E_{\textrm{rec,p}}$ bin, respectively, while $M_{e}$ and $P_{e}$ are the detector-wide measured and predicted content of bin $e$, respectively:
\begin{equation}\label{eq:absSpectrum}
M_{e}=\sum_{l=1}^{6}M_{l,e}
\,\,\textrm{\,and\,}\,\,
P_{e}=\sum_{l=1}^{6}P_{l,e}.
\end{equation}
This form for $\Delta_{l,e}$ is chosen to minimize the dependence of the fitted oscillation parameters on the choice of the input reactor \nuebar model.  
$P_{e}$ was formed by applying the best fit PG4-generated detector response model to IBD interactions following the $^{235}$U \nuebar~energy spectrum of Ref.~\cite{bib:huber} and the cross-section of Ref.~\cite{Vogel:1999zy}.  $P_{l,e}$ was then determined using these inputs, a baseline generator taking into account the finite detector and core sizes, and sterile neutrino oscillation parameters ($\Delta$m$^2_{41}$,sin$^2 2\theta_{14}$) as defined in Eq.~\ref{eq:osc}.  

Statistical and systematic uncertainties and their correlation between energy bins are taken into account through the covariance matrix $V_{\textrm{tot}}$.  
For each systematic uncertainty described in the previous sections, a covariance matrix $V_x$ is produced via generation of toy MC datasets including 1\,$\sigma$ variation of the parameter in question unless otherwise previously specified.  
For signal and background statistical uncertainties, $V_x$ are calculated directly.  
All $V_x$ are then summed to form $V_{\textrm{tot}}$.  

\begin{figure}[h]
\includegraphics[trim = 0.0cm 0.3cm 0.0cm 0.0cm, clip=true, width=0.49\textwidth]{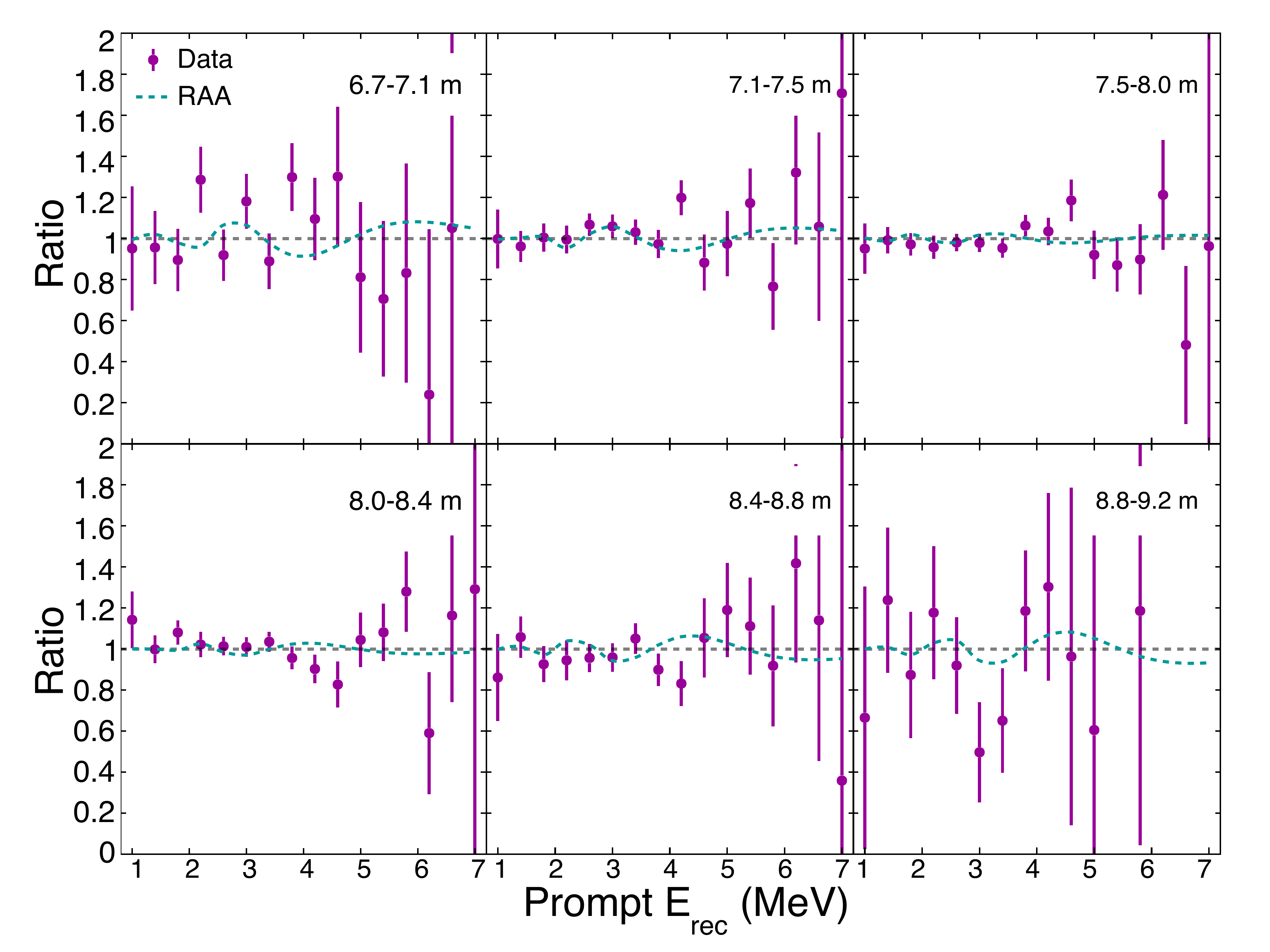}
\caption{Ratio of measured IBD prompt E$_{\textrm{rec},p}$ spectra in six baseline bins from 6.7 to 9.2\,m to the baseline-integrated spectrum.  Also shown are the no-oscillation (flat) expectation and an oscillated expectation corresponding to the the best fit Reactor Antineutrino Anomaly oscillation parameters~\cite{bib:mention2011}.  Error bars indicate statistical and systematic uncertainties, with statistical correlations between numerator and denominator properly taken into account.}
\label{fig:baselines}
\end{figure}

Fig.~\ref{fig:baselines} shows ratios of the measured IBD E$_{\textrm{rec,p}}$ spectra at differing baselines ($M_{l,e}$) to the baseline-integrated measured spectrum ($M_{e}\frac{P_{l,e}}{P_{e}}$).
Also shown are the no-oscillation case (flat line) and the expected behavior due to oscillations matching the best fit parameters of the Reactor Antineutrino Anomaly (dashed line)~\cite{bib:mention2011}. 
No significant deviations from unity are observed at specific baseline or energy ranges.  

\begin{figure}[t]
\includegraphics[trim = 0.0cm 0.0cm 0.0cm 0.0cm, clip=true, width=0.45\textwidth]{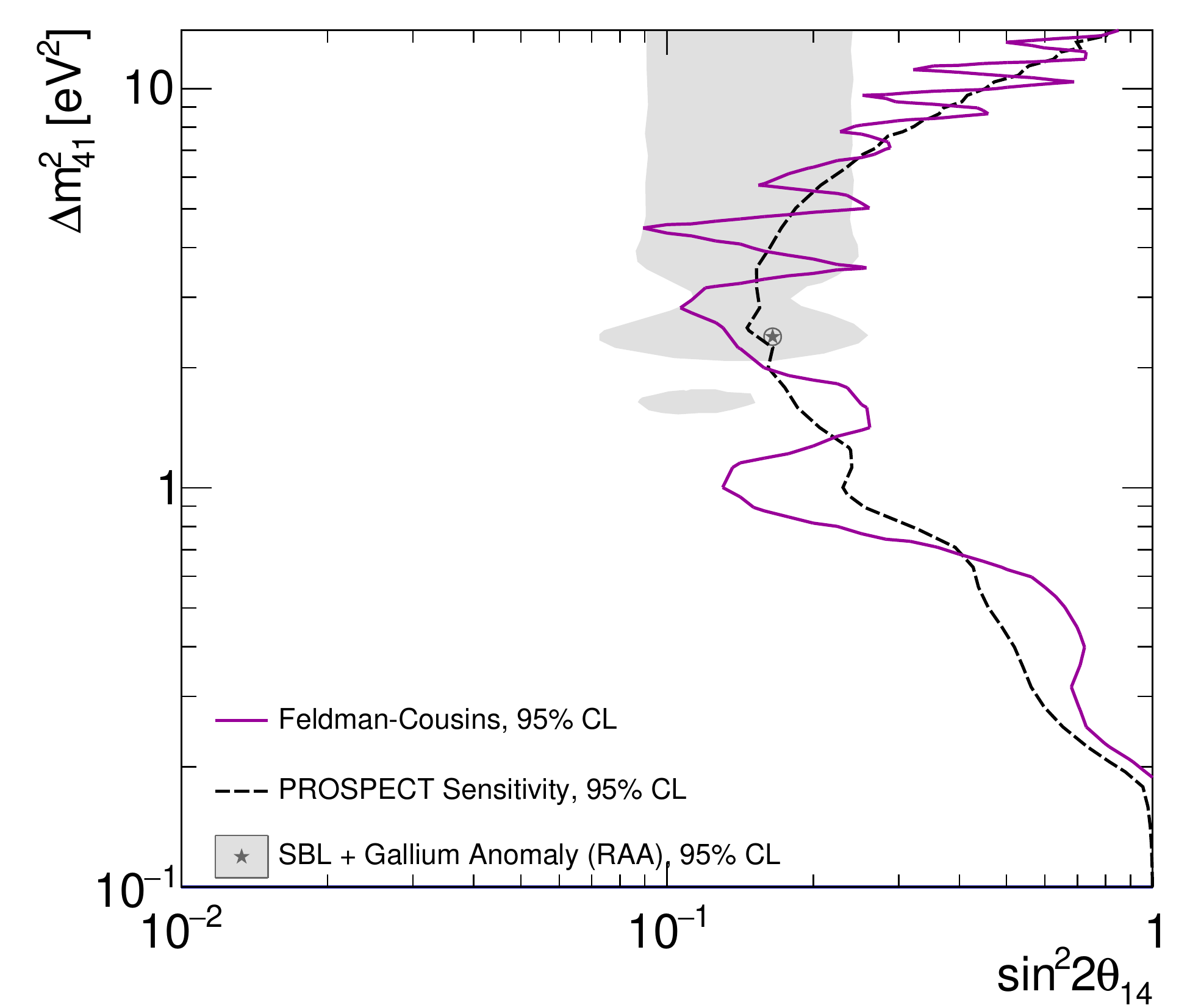}
\caption{Sensitivity and 95\,\% confidence level sterile neutrino oscillation exclusion contour from the 33 live-day PROSPECT reactor-on dataset.  The best fit of the Reactor Antineutrino Anomaly~\cite{bib:mention2011} is disfavored at 2.2\,$\sigma$ confidence level.}
\label{fig:osc}
\end{figure}

This level of agreement is quantified using the $\chi^2$ of Eq.~\ref{eq:abs}.  
At $\theta_{14}$ = 0, the $\chi^2$/NDF is 61.9/80, indicating good agreement between the data and the no-oscillation hypothesis.  
If oscillations are allowed, a global minimum is found at $\Delta$m$^2_{41}$ = 0.5\,eV$^2$ and sin$^2 2\theta_{14}$ = 0.35, with $\chi^2$/NDF = 57.9/78.  
Using a frequentist approach~\cite{fc}, this $\Delta \chi^2$ is found to have an associated p-value of 0.58, indicating little incompatibility with the no-oscillation hypothesis.  
An exclusion contour, shown in Fig.~\ref{fig:osc}, is generated to identify all grid points whose $\Delta \chi^2$ with respect to the best-fit in data exceeds that of 95\,\% (2\,$\sigma$) of oscillated toy datasets generated at that grid point~\footnote{\label{note1}See supplementary material provided with this Letter.}.  
The present dataset excludes significant portions of the Reactor Antineutrino Anomaly allowed region~\cite{bib:mention2011}, and disfavors its best fit point at 2.2$\,\sigma$ confidence level (p-value 0.013).  
The present sensitivity is limited by statistics. 
Shown along with the data exclusion contour is the expected PROSPECT 95\,\% confidence level  sensitivity curve for this dataset.  
This result was further cross checked with an independent oscillation analysis using the Gaussian CLs method~\cite{Qian:2014nha}.  

In summary, the PROSPECT experiment has observed interactions of 25461 reactor $\nuebar$~produced by $^{235}$U fission in 33 live-days of reactor-on running.  
The current signal selection provides a ratio of 1.32 \nuebar detections to cosmogenic backgrounds, as well as the capability to identify reactor-on/off state transitions to 5$\,\sigma$ statistical confidence level within 2\,hours. 
These demonstrate the feasibility of on-surface reactor \nuebar detection and the potential utility of this technology for reactor power monitoring.  
A comparison of measured IBD prompt energy spectra between detector baselines with the 33 live-day dataset provides no indication of sterile neutrino oscillations. 
This disfavors the Reactor Antineutrino Anomaly best fit point at 2.2$\,\sigma$ confidence level and constrains significant portions of the previously allowed parameter space at 95\,\% confidence level. 

This material is based upon work supported by the following sources: US Department of Energy (DOE) Office of Science, Office of High Energy Physics under Award No. DE-SC0016357 and DE-SC0017660 to Yale University, under Award No. DE-SC0017815 to Drexel University, under Award No. DE-SC0008347 to Illinois Institute of Technology, under Award No. DE-SC0016060 to Temple University, under Contract No. DE-SC0012704 to Brookhaven National Laboratory, and under Work Proposal Number  SCW1504 to Lawrence Livermore National Laboratory. This work was performed under the auspices of the U.S. Department of Energy by Lawrence Livermore National Laboratory under Contract DE-AC52-07NA27344 and by Oak Ridge National Laboratory under Contract DE-AC05-00OR22725. Additional funding for the experiment was provided by the Heising-Simons Foundation under Award No. \#2016-117 to Yale University. 

J.G. is supported through the NSF Graduate Research Fellowship Program and A.C. performed work under appointment to the Nuclear Nonproliferation International Safeguards Fellowship Program sponsored by the National Nuclear Security Administration’s Office of International Nuclear Safeguards (NA-241). This work was also supported by the Canada  First  Research  Excellence  Fund  (CFREF), and the Natural Sciences and Engineering Research Council of Canada (NSERC) Discovery  program under grant \#RGPIN-418579, and Province of Ontario.

We further acknowledge support from Yale University, the Illinois Institute of Technology, Temple University, Brookhaven National Laboratory, the Lawrence Livermore National Laboratory LDRD program, the National Institute of Standards and Technology, and Oak Ridge National Laboratory. We gratefully acknowledge the support and hospitality of the High Flux Isotope Reactor and Oak Ridge National Laboratory, managed by UT-Battelle for the U.S. Department of Energy.

\bibliographystyle{apsrev4-1}
\bibliography{refs}{}

\begin{thebibliography}{42}
\bibitem [{dis()}]{disclaimer}%
  \BibitemOpen
  \href@noop {} {}\bibinfo {note} {Certain trade names and company products are
  mentioned in the text or identified in illustrations in order to adequately
  specify the experimental procedure and equipment used. In no case does such
  identification imply recommendation or endorsement by the National Institute
  of Standards and Technology, nor does it imply that the products are
  necessarily the best available for the purpose.}\BibitemShut {Stop}%
\end{thebibliography}

\end{document}